\documentclass[aps,prb,two column,superscriptaddress]{revtex4-1}
\usepackage{graphicx}
\usepackage{amssymb, amsmath}

\newcommand{\LCMO}{$\mathrm{Lu_2}\mathrm{Co}\mathrm{Mn}\mathrm{O}_6$}
\newcommand{\LMCO}{$\mathrm{Lu_2}\mathrm{Mn}\mathrm{Co}\mathrm{O}_6$}
\newcommand{\CCMO}{$\mathrm{Ca_3}\mathrm{Co}\mathrm{Mn}\mathrm{O}_6$}

\newcommand{\Co}{$\mathrm{Co}^{4+}$}
\newcommand{\Mn}{$\mathrm{Mn}^{2+}$}

\newcommand{\ic}{$c$}
\newcommand{\ib}{$b$}
\newcommand{\ia}{$a$}

\begin{document}

\title{Electric polarization observed in single crystals of multiferroic \LMCO\ }


\author{S. Chikara}
\affiliation{National High Magnetic Field Laboratory, Los Alamos National Laboratory, Los Alamos, NM 87545, USA}

\author{J. Singleton}
\affiliation{National High Magnetic Field Laboratory, Los Alamos National Laboratory, Los Alamos, NM 87545, USA}

\author{J. Bowlan}
\affiliation{Center for Integrated Nanotechnologies, Los Alamos National Laboratory, Los Alamos, NM 87545, USA}

\author{D. A. Yarotski}
\affiliation{Center for Integrated Nanotechnologies, Los Alamos National Laboratory, Los Alamos, NM 87545, USA}

\author{N. Lee}
\affiliation{Department of Physics and IPAP, Yonsei University, Seodaemun-gu, Seoul, 120-749, South Korea}

\author{H. Y. Choi}
\affiliation{Department of Physics and IPAP, Yonsei University, Seodaemun-gu, Seoul, 120-749, South Korea}

\author{Y. J. Choi}
\affiliation{Department of Physics and IPAP, Yonsei University, Seodaemun-gu, Seoul, 120-749, South Korea}

\author{V. S. Zapf}
\affiliation{National High Magnetic Field Laboratory, Los Alamos National Laboratory, Los Alamos, NM 87545, USA}

\email[]{schikara@lanl.gov}


\begin{abstract}

We report electric polarization and magnetization measurements in single crystals of double perovskite \LMCO\ using pulsed magnetic fields and optical second harmonic generation (SHG) in DC magnetic fields. We observe well-resolved magnetic field-induced changes in the electric polarization in single crystals and thereby resolve the question about whether multiferroic behavior is intrinsic to these materials or an extrinsic feature of polycrystals. We find electric polarization along the crystalline {\bf b}-axis, that is suppressed by applying a magnetic fields along {\bf c}-axis and advance a model for the origin of magnetoelectric coupling. We furthermore map the phase diagram using both capacitance and electric polarization to identify regions of ordering and regions of magnetoelectric hysteresis. This compound is a rare example of coupled hysteretic behavior in the magnetic and electric properties. The ferromagnetic-like magnetic hysteresis loop that couples to hysteretic electric polarization can be attributed not to ordinary ferromagnetic domains, but to the rich physics of magnetic frustration of Ising-like spins in the axial next-nearest neighbor interaction model.

\end{abstract}

 


\maketitle

\section{Introduction}

\begin{figure*}
\includegraphics[scale=0.6,angle= 0]{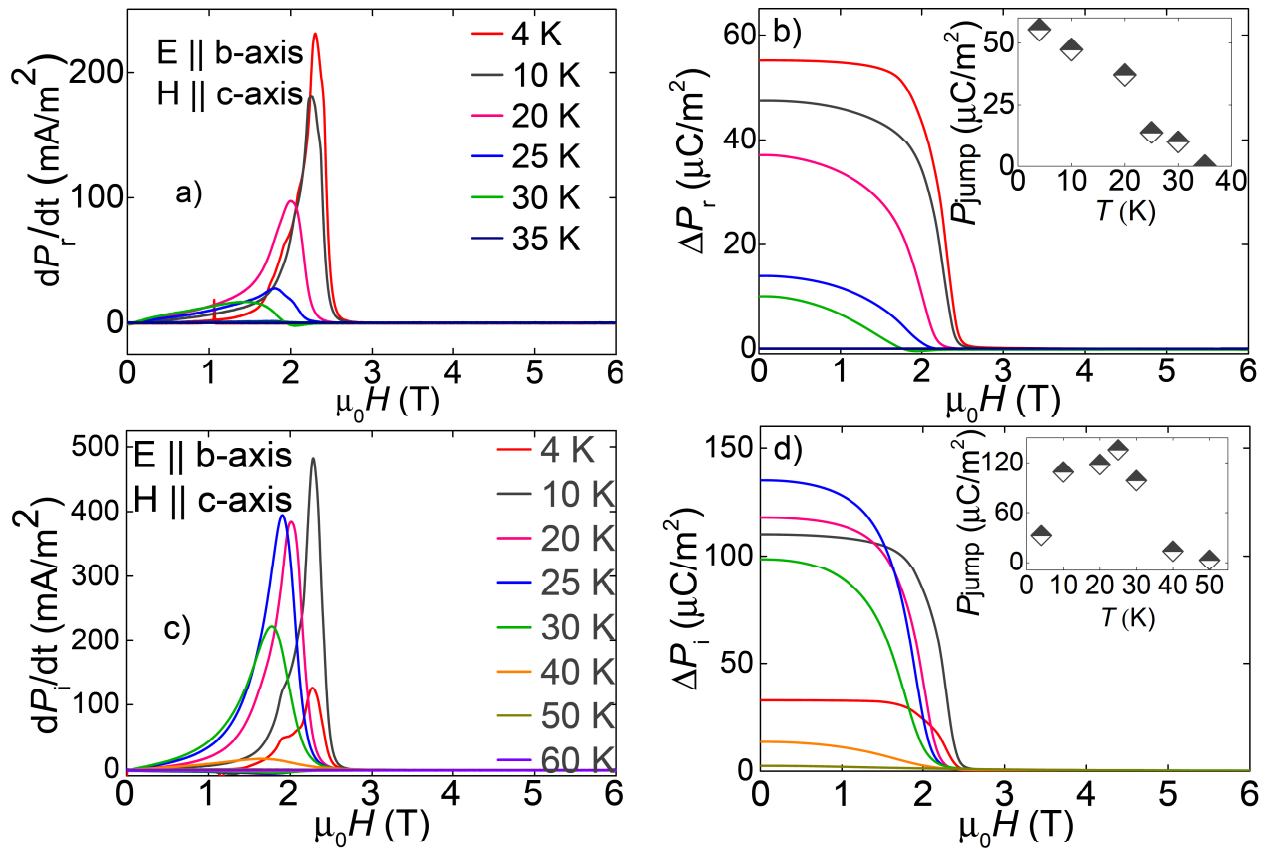}
\caption{The top two figures show the remanent electric polarization $\Delta P_{\rm_r}(H) = P_{\rm_r}(H) - P_{\rm_r}(0)$ measured in $E = 0$ in pulsed magnetic field $H$ for different temperatures $T$. Sample was previously poled by cooling through $T_{\rm N}$ in an applied $E$ and $H = 0$. (a) Shows the raw data ${\rm d}P_r/{\rm d}t$ and (b) the integrated $\Delta P_{\rm r}(H)$. The bottom two figures show the induced electric polarization $\Delta P_{\rm i}(H)$ measured with $E$ applied during the measurements in pulsed magnetic fields $H$ for different temperatures $T$. Sample was previously cooled through $T_{\rm N}$ in $E = H = 0$. (c) shows the raw data ${\rm d}P_i (H)/{\rm d}t$ and (d) the integrated signal $\Delta P_{\rm i}(H)$. The insets shows the magnitude of the magnetic field-induced jump $P_{\rm jump}$ between 0 and 6 T in $\Delta P_{\rm r}(H)$ and $\Delta P_{\rm i}(H)$ for different $T$. All data are for $\textbf{H} \parallel \textbf{c}$ and $\textbf{E} \parallel \textbf{b}$.
}
\label{polarization} 
\end{figure*} 

\noindent 

Multiferroics are materials that exhibit more than one primary ferroic order parameter 
in a single phase.~\cite{Spaldin05, Khomskii06, Khomskii09} 
The most commonly studied combination is ferromagnetic and ferroelectric (FE)
order, which can in turn give rise to magnetoelectric coupling.~\cite{Eerenstein06, Cheong07, Bibes08} 
Lu$_2$MnCoO$_6$ is a double perovskite material (also reported as \LCMO\ in the literature \cite{Lee14}) in which magnetic order breaks spatial inversion symmetry and thereby induces ferroelectricity. Usually in single-phase materials, magnetic order that induces electric polarization 
involves complex, often non-collinear structures that have little or no net magnetization. Unlike such materials, \LMCO\ is a rare example of a bulk multiferroic that exhibits strong,
coupled ferromagnetic-like magnetization $M$ and ferroelectric-like electric
polarization $P$, both of which show hysteresis 
in response to changing magnetic field $H$ and/or temperature $T$.~\cite{YanezVilar11}

\LMCO\ crystallizes in a $\mathrm{P2}_{1}/\mathrm{n}$ space group, in a double perovskite structure with a slight monoclinic distortion.~\cite{YanezVilar11,Lee14}
The \Co and \Mn ions alternate along the ${\bf c}$-axis in distorted 
corner-sharing oxygen octahedra.  
This system orders antiferromagnetically at $T_{\rm N} = 48$~K and 
below a second temperature, $T_{\rm H} = 35$~K,
there is the onset of the above-mentioned 
magnetic and electric-polarization hysteresis with applied magnetic field, plus
strongly frequency-dependent magnetic and electric properties.\cite{YanezVilar11, Zapf14}
Elastic neutron-scattering measurements identify magnetic 
order with a net microscopic magnetization $M = 0$ at $H = 0$, 
despite the observed  magnetic hysteresis loop that develops in 
applied $H$.~\cite{YanezVilar11} 
The magnetic ordering consists of ${\bf c}$-axis chains of Co($\uparrow$)- Mn($\uparrow$)- Co($\downarrow$)- Mn($\downarrow$) or Mn($\uparrow$)- Co($\uparrow$)- Mn($\downarrow$)- Co($\downarrow$),~\cite{YanezVilar11,Zapf14} 
which are consistent with ab-initio calculations.~\cite{Xin2015}  
Magnetization measurements on single crystals demonstrate significant anisotropy in magnetic properties with hysteresis occuring only for 
$H$ along the easy axis,~\cite{Lee14} indicating Ising-like behavior. 
The orbital occupations of Co$^{2+}$ and Mn$^{4+}$ suggest Co$^{2+}$
as the likely source of most of the anisotropy.~\cite{Kim14} 

Muon-spin rotation ($\mu$-SR), elastic neutron diffraction 
and bulk property measurements of \LMCO\ show consistency~\cite{Zapf14} 
with a variant of the axial next-nearest neighbor interaction (ANNNI) model.~\cite{Bak82,Selke88} 
In the ANNNI model, frustration between nearest and 
next-nearest neighbor Ising-like spins gives rise to a very rich phase 
diagram with hundreds of micro-phase transitions between states 
with different long-wavelength incommensurations, and ultimately 
an $\uparrow \uparrow \downarrow \downarrow$ ground state 
at $T = 0$ materializes for certain ratios of exchange interactions. 
Besides the rich spectrum of phases, there is a continuum of low-lying 
excitation states that can generate the hysteresis and dynamics 
observed in this system. In the ANNNI picture, the transition 
from reversible to hysteretic behavior would correspond to the 
freezing of spin-flip excitations in 
the $\uparrow \uparrow \downarrow \downarrow$ state. 
These frustrated magnetic interactions can give rise to magnetic
hysteresis in a system that is not a simple ferromagnet with domains;
such a system can have the necessary low symmetry to sustain electric polarization.
The $\uparrow \uparrow \downarrow \downarrow$ ground state has also previously been observed in the compound Ca$_3$CoMnO$_6$, which produces an electric 
polarization along the $\bf c$-axis due to symmetric exchange striction.~\cite{Choi08,Jo09,Kim14} However, in \LMCO, we will show that the electric polarization 
points along the {${\bf b}$}-axis.    
While coupled electric polarization and magnetization were observed in polycrystalline samples of \LMCO\ for both $\textbf{H} \parallel \textbf{c}$ and $\textbf{H} \perp c$,~\cite{YanezVilar11} previous single-crystal studies were unable to detect electric 
polarization.~\cite{Lee14} 
The single-crystal work did show a peak in the 
dielectric constant near $T = 35$~K for electric fields along the {${\bf b}$}-axis 
that is suppressed by magnetic fields applied along the {${\bf c}$}-axis. 
Here, we report both electric polarization and dielectric measurements 
as a function of magnetic field in single crystals of \LMCO.  
This establishes that the multiferroic behavior is an intrinsic effect and not due to grain boundaries or other extrinsic effects of polycrystals.~\cite{YanezVilar11}  


\section{Experimental Details}
The single crystals were grown using a flux technique.~\cite{Lee14} 
Contacts for capacitance and electric polarization measurement were made by sputtering platinum and attaching leads with silver epoxy.
Magnetic field pulses of 10 and 60~T with a sweep rate $\mu_0 {\rm d}H/{\rm d}t$
of up to  11.62~kT/sec for the upsweep of the pulse were applied using a capacitatively-driven 
resistive 65 T magnet.~\cite{Singleton04}
The magnetoelectric current ${\rm d}P/{\rm d}t$, is generated as charges 
are drawn from ground to the sample contacts to screen the sample's 
surface charge as it changes with magnetic field.~\cite{Zapf10,YanezVilar11} 
The change in electric polarization $\Delta P = P(H) - P(H=0)$ is obtained by numerically integrating ${\rm d}P/{\rm d}t$. The magnetoelectric current in 
response to the time-varying magnetic field was recorded with a Stanford Research 570 current-to-voltage converter at a data acquisition rate of 1 MHz. The resolution of these measurements increases with the square root of ${\rm d}H/{\rm d}t$ and can exceed that of comparable 
superconducting-magnet measurements with 
$\mu_0 {\rm d}H/{\rm d}t \approx 0.01$~T/s 
by three orders of magnitude.~\cite{Zapf10,Zapf11,Kim14} 
Data shown in the figures were taken with an electric field of 
$3.8 \times 10^5$~V/m (maximum electric field that can be used with $^4\rm{He}$) applied before or during the measurement. 
Different electric fields up to this maximum value were applied 
and a linear dependence of the electric polarization on electric field was observed, with a small offset attributable to stray electric fields due to schottky barriers at contacts and thermoelectric voltages. 

Magnetization was measured using pulsed-field extraction 
magnetometry.~\cite{Detwiler2000,Goddard2008a} 
The sample was cooled by immersion in $^3$He or $^4$He and the 
temperature recorded approximately 1~s before the magnetic field pulse. 
The magnetic field is determined by integrating the voltage induced in a coil
close to the field center, and calibrated by observing the de 
Haas-van Alphen oscillations of copper.~\cite{Goddard2008a} 

\begin{figure}
\includegraphics[scale=1.0 ,angle= 0]{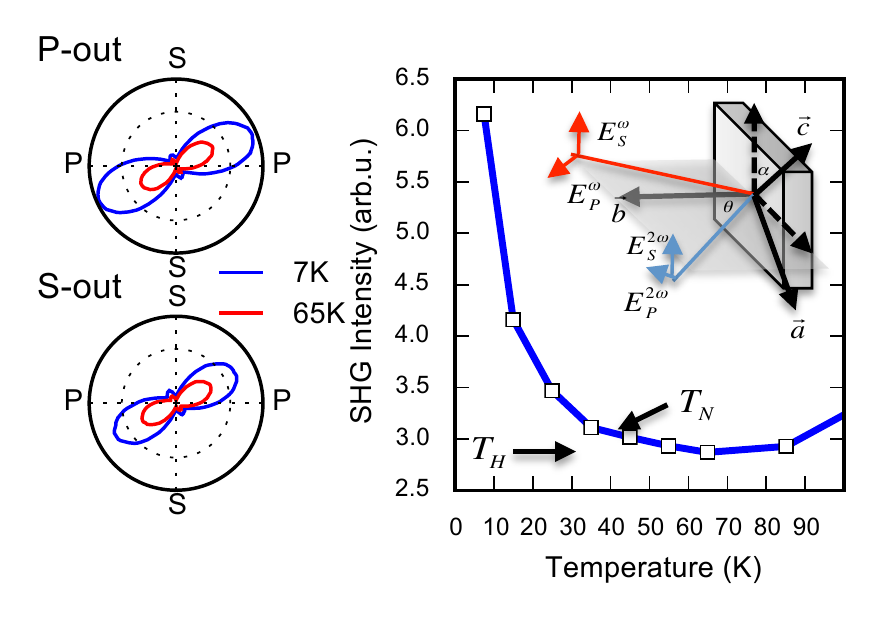}
\caption{The optical SHG measurements on \LMCO\ single crystal. The figure on the left shows the P-out and S-out polarization component. The change in the symmetry between 7 and 65 K is apparent. The right panel shows the SHG intensity as a function of temperature with rapid increase in intensity below 35 K. $E^{2\omega}_S$ and $E^{2\omega}_P$ are the S and P components of the second harmonic polarization respectively. The inset shows the measurement geometry.}
\label{lu2comno6_shg}
\end{figure}

Optical Second Harmonic Generation (SHG) was measured at $H=0$. The measurement geometry is shown in Fig. \ref{lu2comno6_shg}. A 800 nm, 100 fs laser pulse is incident on the \LMCO\ crystal at approximately 45$^{\circ}$. The 400~nm SHG signal is detected with a Photo Multiplier Tube using a lock-in amplifier. The 800 nm light is rejected by a 10 nm interference filter passing 400 nm, and a BG40 colored glass filter. The output polarization, {\bf E} parallel to the plane of incidence (P-out) / {\bf E} perpendicular to the plane of incidence (S-out) is selected with a Glan-Taylor polarizer. We measured the input polarization dependence of the S- and P- polarized components of the SHG signal from 5 to 200 K. The crystal was dry-polished to a roughness 250 nm with diamond paper. The {\bf b}-axis is normal to the sample surface, and the {\bf a}- and {\bf c}- axes are rotated 45$^{\circ}$ to the plane of incidence.


\section {Experimental Results and Discussion}

The electric polarization was measured while sweeping the magnetic field in two different ways. First by poling (cooling through $T_{\rm N}$ in an electric field $E$ at $H = 0$) 
and then setting the applied $E$ to zero and measuring the change in polarization, $\Delta P_r (H)$.
In the  other set of measurements, the sample is first cooled through $T_{\rm N}$ with $E = H = 0$. Next, $E = 3.8 \times 10^5$ V/m is applied; the magnetic field is pulsed while the electric polarization change, $\Delta P_{\rm i}(H)$ is recorded. These data measures field-induced capacitance changes, 
$C(H)-C(H=0) = 1/E[P_{i}(H)-P_{i}(0)]$.

Figures \ref{polarization} (a) and (b) show the remanent electric polarization as a function of magnetic field $\Delta P_r (H) = P_r(H) - P_r(0)$, and the raw ${\rm d}P_r (H)/{\rm d}t$ data from which it was derived. The data are shown for $\textbf{E} \parallel \textbf{b}$ and $\textbf{H} \parallel \textbf{c}$. All other combinations of $\textbf{E}$ and $\textbf{H}$ along the {${\bf a}$}, {${\bf b}$}, {${\bf c}$}-axes were also measured; however, no electric polarization was detected using these combinations. 
The observation of $\Delta P_{\rm r}(H)$ only along $\textbf{E} \parallel \textbf{b}$ for $\textbf{H} \parallel \textbf{c}$  is consistent with the single-crystal dielectric measurements in applied magnetic field reported earlier.~\cite{Lee14} 
These data at first sight seem   inconsistent with the polycrystal result that finds $\Delta P_{\rm r}(H)$ for both $\textbf{H} \parallel \textbf{P}$ and $\textbf{H} \perp \textbf{P}$.
However, the latter result could be ascribed to magnetic dipole interactions
between grains of the polycrystals.~\cite{YanezVilar11} 

$\Delta P_{\rm r}(H)$ measured in the single crystals of the current study is an order of magnitude larger than what was observed in polycrystalline samples, which is to be expected for a uniaxial electric polarization.~\cite{Lee14,YanezVilar11}  
In Fig. \ref{polarization} (a) and (b), there is an onset of remanent electric 
polarization $\Delta P_{\rm r}(H)$ just below $T_{\rm H} = 35$~K.
Similarly, there is an onset of magnetic hysteresis at this temperature in the $M(H)$ data, 
also seen in previous poly- and single-crystal measurements.~\cite{YanezVilar11,Lee2014} $\Delta P_{\rm r}(H)$ is suppressed in magnetic fields above $\approx 2$~T, which corresponds to the coercive magnetic field in the $M(H)$ hysteresis loop for ${\bf H} || {\bf c}$. 
This $H$ and $T$ dependence of $\Delta P_r$ is consistent with the previous polycrystalline results.~\cite{YanezVilar11}

On the other hand, Figures~\ref{polarization} (c) and (d) show the 
electric polarization $\Delta P_{\rm i}(H)$ that is induced by a DC electric field that is applied during the magnetic field pulse. A non-zero change in $P_{\rm i}$ with $H$ is observed for 
$T < T_{\rm N} = 50$~K, rather than below $T_{\rm H} = 35$~K as was seen for $\Delta P_{\rm r}$. 
The dielectric response $\Delta P_{\rm i}(H)$ does show a peak near $T_{\rm H} = 35$~K, however, which is consistent with the previous capacitance measurements
of single- and polycrystal samples in
quasistatic superconducting magnets.~\cite{YanezVilar11,Lee14} The magnitude of our measured $\Delta P_{\rm i}(H)$ is three times larger than our measured $\Delta P_{\rm r}(H)$ at 4 K.

\begin{figure}
\includegraphics[scale=0.25,angle= 0]{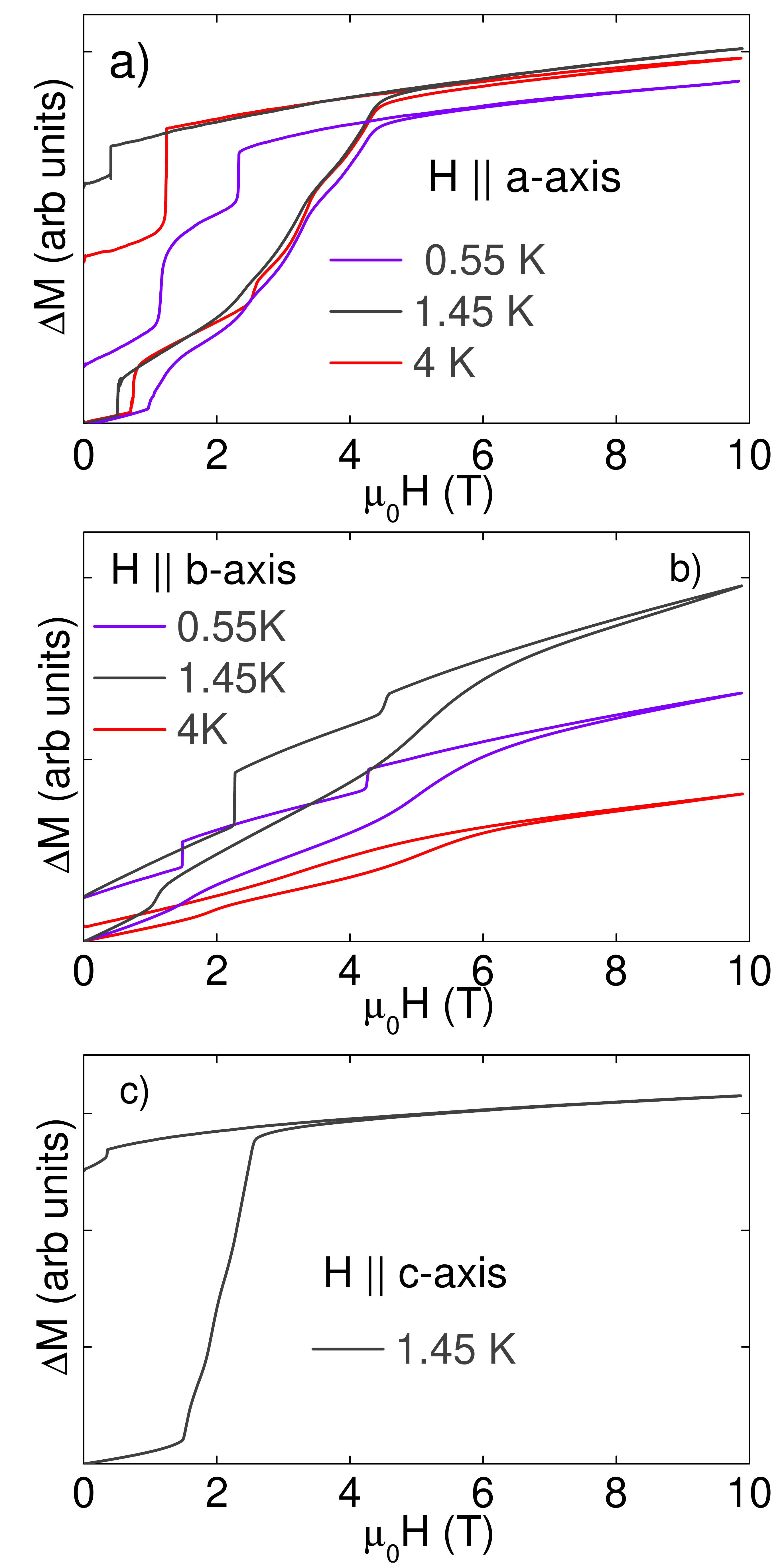}
\caption{Magnetization change $\Delta M = M(H) - M(H=0)$ vs magnetic field $H$ in pulsed magnetic fields for $H$ along the a) \ia-axis, b) \ib-axis and c) \ic-axis. The pulsed data along different axes are not to scale - note the DC magnetization data show the $c$-axis magnetization to be significantly larger than along $a$ or $b$.~\cite{Lee14}}
\label{MagnetizationB} 
\end{figure}

\begin{figure}
\includegraphics[scale=0.85,angle= 0]{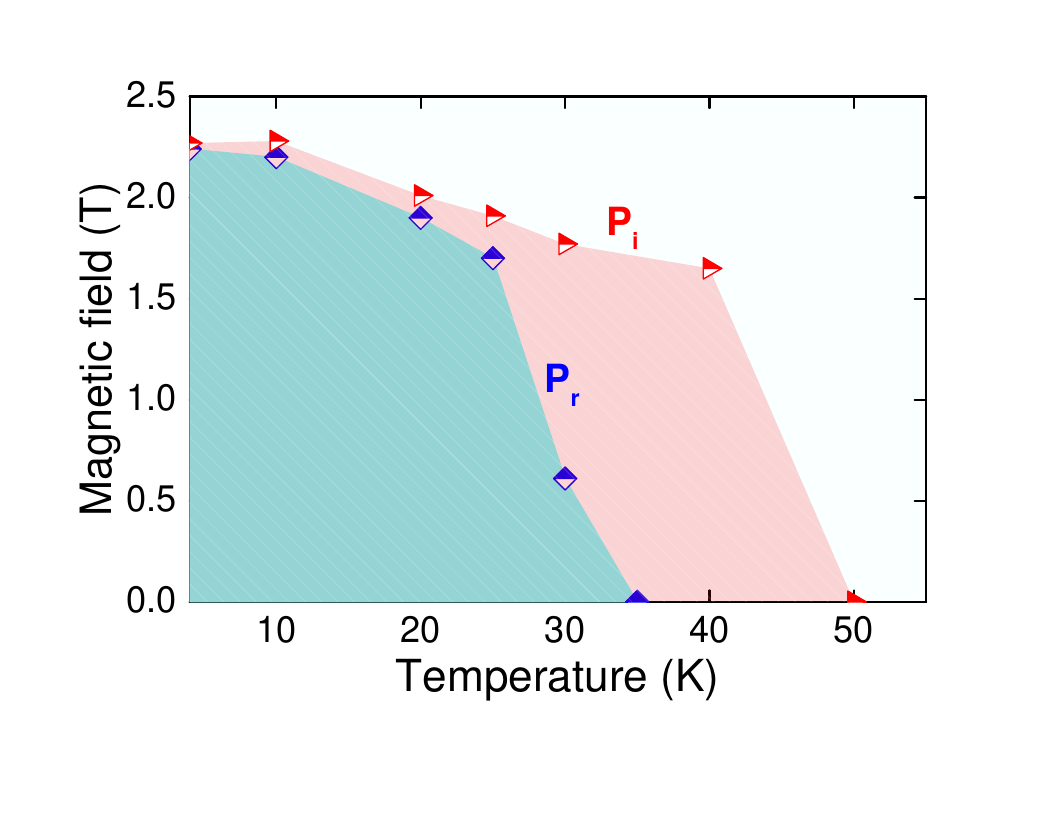}
\caption{The magnetic field vs temperature phase diagram showing the region where 
the remanent polarization $\Delta P_{\rm_r}$ and the induced polarization $\Delta P_{\rm_i}$ are non-zero. The region of non-zero $\Delta P_{\rm_i}$ corresponds to magnetic ordering in thermodynamic neutron diffraction measurements while $\Delta P_{\rm_r}$ corresponds to magnetic and electric hysteresis.}
\label{phase_diagram}
\end{figure}

\begin{figure}
\includegraphics[scale=0.3,angle= 0]{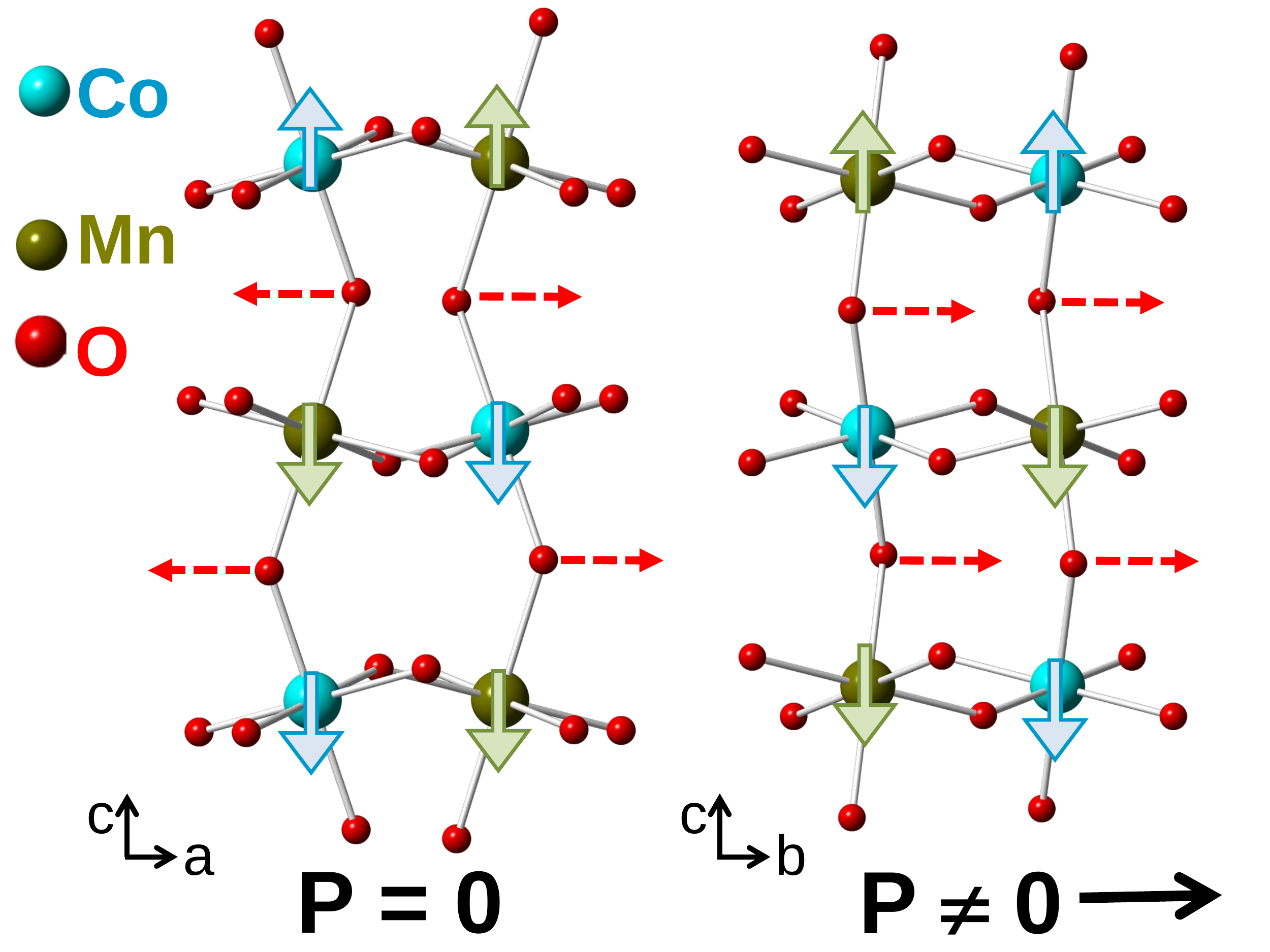}
\caption{The projection of \LCMO\ structure in the (a)~$ac$ and (b)~$bc$ planes. The red arrows indicate the direction of oxygen displacement in response to the bond shortening or elongation due to the exchange between parallel or antiparallel spins on neighboring ions.}
\label{strr_projections}
\end{figure}

We show the magnetization in pulsed magnetic fields $\Delta M(H) = M(H)-M(H=0)$ for $\textbf{H}$ along the {${\bf a}$}, {${\bf b}$} and {${\bf c}$}-axes in Fig.~\ref{MagnetizationB}. $\Delta M(H)$ for $\textbf{H} \parallel \textbf{c}$ is hysteretic in field and consistent with previous results in pulsed and DC fields for single- and polycrystals.~\cite{YanezVilar11,Lee14} The coercive magnetic field is 2.0~T at 1.5~K. On the other hand, $\Delta M(H)$ for $\textbf{H} \parallel \textbf{a}$ and $\textbf{H} \parallel \textbf{b}$ differs from previous single crystal results in DC fields. $\Delta M(H)$ in pulsed fields shows hysteresis and step-like behavior while $M(H)$ in DC magnets is smooth, continuous, and non-hysteretic. Similar steps in the magnetization, electric polarization and magnetostriction were observed in \CCMO\ induced by certain ranges of magnetic field sweep rates between 75 and 1,500~T/s while smooth behavior was observed for both faster and slower sweep rates.~\cite{Kim15} 
In \LMCO\, $P(H)$ is zero for the field orientations   where steps are seen 
in pulsed-field $\Delta M(H)$ to 1700~T/s.~\cite{Chikara15b} In similar systems like \CCMO\ and Ca$_3$Co$_2$O$_6$, many theories have been advanced for the origin of the metastable steps, most recently the presence of excited metastable states with high degeneracy due to frustration in Ising-like spins.~\cite{Kamiya12a}

We measured the optical second harmonic generation (SHG) in reflection from the single crystal of \LMCO.
 The $\bf b$-axis is normal to the sample surface, and the $\bf a$ and $\bf c$ axes are rotated 45 $^\circ$ to the plane of incidence. A small SHG signal is observed at all temperatures up to 300 K. This signal most likely originates from the crystal surface where inversion symmetry is broken.~\cite{Denev11,Fiebig2005}  The SHG signal amplitude is essentially flat down to ~40~K, and below 40 K the SHG signal begins to rise rapidly indicating that inversion symmetry has been broken.
This pattern is consistent with the behavior of paraelectric / ferroelectric phase transitions in many other materials, ~\cite{Denev11} and is strong evidence that \LMCO\ is ferroelectric below the temperature, $T_{\rm  H}$ = 35 K, where hysteretic polarization appears. 

Figure \ref{phase_diagram} shows the $(T,H)$ phase diagram of \LMCO\ from the $\Delta P(H)$ and $\Delta M(H)$ measurements performed in pulsed magnetic fields. The remanent electric polarization $\Delta P_{\rm i}(H)$ is nonzero below $T =35$~K, whereas the polarization induced by applied electric fields during the measurement $\Delta P_{\rm i}(H)$  is nonzero below a higher temperature of 50~K. This implies that the capacitance shows an enhancement and a magnetic field dependence below the magnetic ordering temperature of 50~K. On the other hand, the hysteresis in the electric polarization, measured by $\Delta P_{\rm r}(H)$, becomes finite below the temperature where hysteresis in $T$ and $H$ turns on in the magnetization at 35~K.~\cite{Zapf14,YanezVilar11}

\LMCO\ shows many similarities to \CCMO\, which also contains chains of alternating Mn$^{4+}$ and Co$^{2+}$ spins, although the oxygen cages have different symmetry.~\cite{YanezVilar11, Kaushik11} In that compound the magnetic ground state is also $\uparrow \uparrow \downarrow \downarrow$ along ${\bf c}$-axis chains. The important difference is that in \CCMO\ $P(H)$ is observed along the {${\bf c}$}-axis, while in \LMCO\ $P(H)$ is seen for $\textbf{P} \parallel \textbf{b}$ with $\textbf{H} \parallel \textbf{c}$. The two compounds have different crystal structures; \CCMO\ is not a double perovskite. In \LMCO\ we can understand the electric polarization along the {\bf c}-axis due to the presence of oxygen bonds all canted along {${\bf b}$} direction.~\cite{Xin2015}   

Figure \ref{strr_projections} (a) and (b) shows the crystal structure of \LMCO\ 
in the $bc$ and $ac$ planes. If $P$ results from an exchange-striction mechanism in the magnetically ordered phase, we can expect that the bonds between ferromagnetically-aligned spins contract while antiferromagnetically aligned spins repel. (Even if the opposite is the case i.e. [ferromagnetically-aligned spins repel while antiferromagnetically-aligned spins attract], the conclusion will be the same). When exchange striction distorts the bonds, the displacements of negatively-charged oxygens relative to positively-charged Mn and Co ions are indicated with arrows. It can be seen that along the $b$-axis, the net electric polarization change due to oxygen displacement in bonds subject to exchange striction on neighboring chains is non-zero while along all other axes, the oxygen displacements on neighboring chains cancel each other out. A net $P$ along $\bf b$ was also predicted for isostructural Y$_2$MnCoO$_6$~\cite{Xin2015} in an $\uparrow \uparrow \downarrow \downarrow$ magnetic ground state scenario for similar reasons. The magnetic field-induced suppression of $P$ occurs abruptly near the saturation magnetization where $\uparrow \uparrow \downarrow \downarrow$ state transitions to $\uparrow \uparrow \uparrow \uparrow$.

In conclusion, we demonstrate electric polarization in single crystals of \LMCO\ for the first time, by using a sensitive pulsed-magnetic-field and optical second-harmonic-generation techniques. We find that \LMCO\ is polar along the ${\bf b}$-, and not along the ${\bf c}$-axis. We show that a dielectric response emerges below the magnetic ordering temperature $T_{\rm N}$ while a hysteretic electric response occurs in the region of ($T,H$) phase space corresponding to hysteresis in the magnetization. We ascribe the origin of the field-induced electric polarization change to exchange striction coherently displacing oxygen ions along the $\bf b$-axis.We also report additional metastable steps in the magnetization at finite magnetic field sweep rates up to 1700~T/s that do not occur at superconducting magnet sweep rates.

Work at LANL was supported by the Laboratory-Directed Research and Development program (LDRD). The NHMFL Pulsed-Field Facility is funded by the U.S. National Science Foundation through Cooperative Grant No. DMR-1157490, the State of Florida, and the U.S. Department of Energy. Work at Yonsei University was supported by the NRF Grant (NRF-2013R1A1A2058155, NRF-2014S1A2A2028481, and NRF-2015R1C1A1A02037744), and partially by the Yonsei University Future-leading Research Initiative of 2014(2014-22-0123).

\bibliography{library}

\end{document}


\title{Electric polarization observed in single crystals of multiferroic \LMCO\ }


\author{S. Chikara}
\affiliation{National High Magnetic Field Laboratory, Los Alamos National Laboratory, Los Alamos, NM 87545, USA}

\author{J. Singleton}
\affiliation{National High Magnetic Field Laboratory, Los Alamos National Laboratory, Los Alamos, NM 87545, USA}

\author{J. Bowlan}
\affiliation{Center for Integrated Nanotechnologies, Los Alamos National Laboratory, Los Alamos, NM 87545, USA}

\author{Dmitry A. Yarotski}
\affiliation{Center for Integrated Nanotechnologies, Los Alamos National Laboratory, Los Alamos, NM 87545, USA}

\author{N. Lee}
\affiliation{Department of Physics and IPAP, Yonsei University, Seodaemun-gu, Seoul, 120-749, South Korea}

\author{H. Y. Choi}
\affiliation{Department of Physics and IPAP, Yonsei University, Seodaemun-gu, Seoul, 120-749, South Korea}

\author{Y. J. Choi}
\affiliation{Department of Physics and IPAP, Yonsei University, Seodaemun-gu, Seoul, 120-749, South Korea}

\author{V. S. Zapf}
\affiliation{National High Magnetic Field Laboratory, Los Alamos National Laboratory, Los Alamos, NM 87545, USA}

\email[]{schikara@lanl.gov}


\pacs{}
 


\maketitle
\noindent Fig.~\ref{supplement} shows the magnetic-field sweep rate. The right axis shows the derivative of magnetic field which gives the sweep rate. The inset shows the steps seen in magnetization, $\Delta M$ as a function of magnetic field at 0.55 K for $H || a-$axis. The steps are seen up to 5~T with sweep rate ranging from 1.43 kT/sec for a 10 T pulse to 11.62~kT/sec for 60~T pulse (not shown here). \\
\begin{figure}
\includegraphics[scale=0.23,angle= 0]{Figure-supplement}
\caption{Left axis and black plot: magnetic field as a function of time for a 10 T pulesd field shot. Right axis and red plot: the derivative of magnetic field with time. Inset: $\Delta M$ as a function of magnetic field for a 10 T shot at 0.55 K. The step-like behavior survives to about 5~T.
}
\label{supplement} 
\end{figure} 

\begin{figure}
\includegraphics[scale=0.50,angle= 0]{shg_supplement}
\caption{The optical SHG measurements on \LMCO\ single crystal. The figure on the left shows the P-out and S-out polarization component. The change in the symmetry between 7 and 65 K is apparent. The right panel shows the SHG intensity as a function of temperature with rapid increase in intensity below 35 K. $E^{2\omega}_S$ and $E^{2\omega}_P$ are the S and P components of the second harmonic polarization respectively. The inset shows the measurement geometry.}
\label{supplement2} 
\end{figure} 

\noindent Fig.~\ref{supplement2} shows the optical SHG measurements on \LMCO\ single crystals. The right panel shows the SHG intensity as a function of temperature for an extended temperature range. The enhancement in the SHG intensity below 35 K is much larger than the slight increase seen at higher tempertaures. The anomaly around 120~K might be due to surface symmetry breaking effects. Since there are no magnetic signatures observed around 120~K in the magnetization data (see main text), it seems unlikely to be related to multiferroic behavior.
